\documentclass{emulateapj}
\usepackage{natbib}
\usepackage{epstopdf}
\usepackage[colorlinks=true,citecolor=blue,urlcolor=magenta,breaklinks]{hyperref}
\slugcomment{Draft in preparation \today}

\shorttitle{Looking for periodicities in the sunspot time series}
\shortauthors{I. Lopes \& H.G. Silva}
\begin{document}
\title[Looking for granulation and periodicity imprints in the sunspot time series]{Looking for granulation and periodicity imprints in the sunspot time series}

\author{Il\'\i dio Lopes,\altaffilmark{1,2,5} \& Hugo G. Silva\altaffilmark{3,4,5}}
 
\altaffiltext{1}{Centro Multidisciplinar de Astrof\'{\i}sica, 
Instituto Superior T\'ecnico, Universidade de Lisboa, Av. Rovisco Pais, 1049-001 Lisboa, Portugal} 
\altaffiltext{2}{Departamento de F\'\i sica,
Universidade de \'Evora, Col\'egio Ant\'onio Luis
Verney, 7002-554 \'Evora - Portugal} 
\altaffiltext{3}{Departamento de F\'\i sica, 
ECT, Instituto de Ci\^encias da Terra, Universidade de \'Evora, 
Rua Rom\~ao Ramalho 59, 7002-554 \'Evora, Portugal}
\altaffiltext{4}{Atmospheric Chemistry Research Group, 
University of Bristol, Cantock’s Close, Bristol, BS8 1TS, UK} 
\altaffiltext{5}{E-mails: (IL) ilidio.lopes@tecnico.ulisboa.pt; (HGS) hgsilva@uevora.pt} 

 
\begin{abstract}
The sunspot activity is the end result of the cyclic destruction and regeneration of magnetic fields by the dynamo action. We propose a new method to analyze the daily sunspot areas data recorded since 1874. By computing the power spectral density of daily data series using the Mexican hat wavelet, we found a power spectrum with a well-defined shape, characterized by three features.  The first term is the 22 yr solar magnetic cycle, estimated in our work to be of 18.43 yr. The second term is related to the daily volatility of sunspots. This term is most likely produced by the turbulent motions linked to the solar granulation. The last term corresponds to a periodic source associated with the solar magnetic activity, for which the maximum of power spectral density occurs at 22.67 days. This value is part of the 22--27 day periodicity region that shows an above-average intensity in the power spectra. The origin of this 22.67 day periodic process is not clearly identified, and there is a possibility that it can be produced by convective flows inside the star.  The study clearly shows a north-south asymmetry. The 18.43 yr periodical source is correlated between the two hemispheres, but the 22.67 day one is not correlated. It is shown that towards the large timescales an excess occurs in the northern hemisphere, especially near the previous two periodic sources. To further investigate the 22.67 day periodicity we made a Lomb-Scargle spectral analysis. The study suggests that this periodicity is distinct from others found nearby.
\end{abstract}


\section{Introduction}
Our current interest in studying the solar activity follows our concern 
in understanding  the mechanisms responsible for the solar dynamo,
as well as to estimate the impact that the sunspot variability has on the total solar irradiance.
As is commonly understood, the cyclic variation of  magnetic fields generated by the dynamo in the Sun's 
interior leads to the formation of a cyclic pattern of sunspot pairs  at the solar surface.
This pattern, which has been observed in the Sun during the past few centuries, shows
a large variability. The exact mechanism responsible for such behavior remains unknown.
In this work, we investigate the origin of such variability by studying the sunspot area data sets.
These data are known to be one of the best proxies of the magnetic activity
at the surface of the Sun. 
Reviews of sunspot properties and their impact on  solar irradiance can be found in the literature~\citet[][]{rev-Solanki2003,2004A&ARv..12..273F,2010LRSP....7....1H,2010LRSP....7....6P,2013ARA&A..51..311S}. 

\smallskip
This work is also relevant for researchers interested in the Sun-Earth connection. 
The increase of sunspot activity is always accompanied 
by an increase of facula activity. These processes are the main drivers of 
the total solar irradiance variability. From  maximum to minimum of the sunspot cycle
the solar irradiance varies  $0.1\%$. Such small differences
are enhanced by specific mechanisms of the Earth's atmosphere of the oceans,
leading to a significant impact on the Earth's climate~\citep[e.g.,][]{2013ARA&A..51..311S}. 
    
\smallskip    
Since Galileo's times in the seventeen century, the systematic observation and counting of the 
sunspots has been carried out, providing definitive evidence of the dynamics, evolution, and 
longevity of the solar magnetic field.  Following  in the  footsteps of the Royal 
Greenwich Observatory (RGO),  the Solar Physics Team from the NASA's Marshall 
Space Flight Center, in collaboration with other research institutions, recently organized a 
series of tables on the sunspot numbers and areas observed daily on the Sun's surface~\citep{2010LRSP....7....1H}. 
This systematic observational work has been fundamental to study the long-term variations 
of the solar magnetic field. Previous studies have found many periodicities on these data series, 
of which the origin of most is still unknown. Examples are the  quasi-biennial periodicity, 
with a period between 1 and 4 yr~\citep{2002A&A...394..701K,2004IAUS..219..128B},
and the Rieger-type periodicity, with a period of 150 days~\citep{2010ApJ...709..749Z}. 
The first periodicity was suggested to be linked to a second dynamo mechanism~\citep{1998ApJ...509L..49B} 
or to a quadrupolar component of the magnetic dynamo configuration~\citep{2004IAUS..219..128B,2013ApJ...765..100S}.
The  Rieger periodicity has been related to the Rossby waves~\citep{2010ApJ...709..749Z}
or to a sub-harmonic period of a fundamental period of 26 days~\citep{1992ApJ...397..337S}.
These sunspot periodicities have been observed in other solar activity phenomena.
In particular, the  Rieger periodicity originally discovered in the gamma-ray observations~\citep{1984Natur.312..623R} has been observed in other solar data, such as coronal mass ejections~\citep{2003MNRAS.345..809L}. 
A review about what is known on these periodicities can be found in~\citet{2010LRSP....7....1H} and~\citet{2010LRSP....7....6P}.

\smallskip

The periodic evolution of the solar magnetic cycle and some of their consequences can be 
substantially explained within the modern description of the solar dynamo theory. The most 
successful method to compute  the solar dynamo model has been introduced by 
\citet{art-Parker1955}  and~\citet{book-Moffatt1978}, among others. The field has 
progressed  remarkably in the past decade owing to the new data made available by 
helioseismology~\citep{rev-Charbonneau2005,2009AnRFM..41..317M,rev-Charbonneau2010}.  

In a nutshell, the solar dynamo begins at the sunspot minimum with a global dipolar magnetic 
field just above the radiative core that runs inside the Sun from the south Pole to the north Pole. 
The differential rotation shears this dipolar field, which gets stretched as it is wrapped around 
the internal radiative core, within a thin boundary layer, between the radiative interior and the 
convection region, usually known as the tachocline layer. In this thin layer, the toroidal 
component of the solar magnetic field is regenerated and stored~\citep{art-GoughMcIntyre1998}. 
The magnetic field being produced in the tachocline increases significantly in intensity, by
stretching and wrapping around the radiative globe. Eventually, it rearranges itself to form tubes 
of magnetic flux that become strong enough to become buoyant, rise to the surface, and break 
through it, in active region belts of  several bipolar sunspot pairs.

\smallskip

This pictorial description of the Sun is quite realistic, and it has been validated by several 
numerical simulations~\citep{art-MJaramilloNandyMartens2009,art-GuerreroDG2009,art-JiangChatterjeeChoudhur2007,
art-DikpatiGilman2006,art-DikpatiCharbonneau1999}.
These hydromagnetic dynamo models are computed in the regime known as the kinematic approximation, 
which consists in considering that the magnetic field is transported by the internal flows.
This class of dynamo models, usually known as flux transport dynamo (FTD), 
has an advantage over other types, as the leading motion flows are inputs of the  model. 
FTD models are the preferred modeling framework to
study the spatiotemporal evolution of the large-scale magnetic field on timescales 
spanning from a few months to several centuries.  Their two primary defining features are 
(i) the observed equatorward migration of sunspot source regions and poleward migration of surface fields,  
both of which are driven by the conveyor belt action of the meridional flow;
(ii) and the period cycle is the primary quantity regulating the speed of the meridional flow.
These models have been quite successful in explaining the main features of the long-term magnetic variability. However, some important physical processes are ignored, such as the back-reaction of the  magnetic field in the motion of the plasma, i.e., the effect of the Lorentz force~\citep{Tobias1997}; even so, the  qualitative results obtained are well consolidated.
\smallskip

Sunspot numbers are good indicators of the Sun's magnetic activity, although more precise 
information about the Sun's magnetic field activity is contained in the sunspot area data 
concerning the two Sun's hemispheres~\citep[e.g.,][]{2010SSRv..155..371P,20014Ilopes}.
In an attempt to unravel the physical mechanisms 
responsible for the formation and evolution of the magnetic field, i.e., the sunspots observed on the Sun's surface, 
we have made a fresh analysis of the sunspot area data publicly available. 
In particular, we discussed the possibility that the granulation in its different forms 
could be a source of the localized power excess observed in the data series. Furthermore, 
we tested our hypothesis by means of a realistic artificial data series, where we investigated the 
different sources of excitation provided.

\smallskip
In this work, we address the impact that the granulation has 
on the formation and evolution of the surface magnetic fields and
attempt to quantify their contribution for the evolution of the magnetic activity over
several centuries. This research complements other recent
work in the topic~\citep[e.g.,][]{2010ApJ...725.1082H,2010cosp...38.1770K,2014ApJ...785...90R}.    
In particular, we clearly find a stochastic component on the sunspot area time series, which
very likely is associated with the random nature of sunspot areas modified by the turbulent
processes produced by the ascendant convective flows. 
Moreover, we propose a method of analysis to disentangle the different components of the 
sunspot area signal, namely, the patterns produced by different mechanisms that contribute 
to the evolution of the magnetic field. We tested the method with realistic artificial data for the 
sunspot area time series. Our analysis of these data shows clearly the existence of two 
components, which we distinguish as the long-term magnetic cycle reversal and the stochastic 
noise of granulation created by the eddy motion process. 

In section~\ref{sec-helioFTD},  we revise some of the leading characteristics of   
FTD models. In section~\ref{sec-ndsmc}, 
we discuss a new method of data analysis of sunspot areas. The data analysis study 
is presented in section~\ref{sec-obs}. In section~\ref{sec-smc}, we discuss a simple theoretical model to 
interpret the data. A detailed  discussion of observational and theoretical
results is given in section~\ref{sec-discussion}, and our conclusions are presented in section~\ref{sec-Conclusion}.

\section{Flux transport dynamo and helioseismology}
\label{sec-helioFTD}

The descriptive power of FTD  models 
comes from the use of the flow velocity field as an input quantity. During the past decades, the 
velocity flow has progressed from a prescribed theoretical parametrized expression, 
based on some physical  considerations, to velocity flows measured directly in the solar interior 
with great accuracy by helioseismology.
Currently, the leading fluid motions measured by helioseismology    
are the differential rotation for which the solar equator rotates faster than the poles
--a property observed on the surface, as well as at the base of convection zone --
and the meridional circulation, a weak flow of material along the meridian lines  
from the equator toward the poles at the surface and 
in the reverse direction below the surface~\citep[e.g.,][]{art-Howe2009}.

\smallskip
These two ingredient profiles, the angular velocity (differential rotation) 
and the meridional flow, are fundamental to current FTD models.
Although the first ingredient is accurately determined from helioseismology data up to the base
of the convection zone, the meridional flow responsible for the basic cellular structure of the meridional 
flow is only measured in a shallow subsurface layer beneath the photosphere.
A current profile of differential rotation can be found in~\citet{art-Schouetal2002}
and the measurements of the meridional circulation flow in~\citet{art-ZhaoKosovichev200} and~\citet{2013ApJ...774L..29Z}.  
The exact cellular structure of the meridional flow is still open to discussion. 
Some authors suggest that the solar convection zone could have been one-, double- or triple-cell  
of meridional circulation~\citep{2014ApJ...785...49P}.

\smallskip
A picture is emerging for the formation and evolution of sunspots within the FTD models.
Sunspots are formed as the result of the large-scale motions inside the Sun, like
differential rotation~\citep{art-Howe2009} and the meridional flow in the solar convection zone~\citep{2007AN....328..264S}.
The combination of these flows and their interaction with the magnetic field set up by the moving, 
electrically charged particles in the solar plasma by dynamo action is believed to create the observed sunspot cycle~\citep[e.g.,][]{2009AnRFM..41..317M,rev-Charbonneau2010}.

\smallskip
The need for more reliable predictions of the solar cycle opened the way to the development of 
more robust dynamo models. Major progress can be achieved by the inclusion of a more 
detailed description of the velocity field that takes into account the presence of the turbulent flows 
present in the solar interior~\citep{art-Howe2009}. There have been hints that the different types 
of circulation flows in the convection zone can play an important role in the cyclic behavior of 
the solar magnetic cycle, in addition to than the angular velocity and meridional flow, these last ones 
being the most well known~\citep[e.g.][]{art-ZhaoKosovichev2004,2013ApJ...774L..29Z}. 
Recent results from  helioseismology reveal the existence of solar cycle variations of solar rotation
-- motion flows with migrating  bands of slower- and  faster-than-average zonal flow.
The so-called torsional oscillations  present evidence of 
penetrating at different depths of the Sun's interior. Torsional bands of slower rotation might 
penetrate close to the base of the convection zone, while bands of faster rotation appeared 
to reach about 10\% below the solar surface~\citep{1980ApJ...239L..33H,art-Vorontsovetal2002,art-Howeetal2005,art-Howeetal2006}, 
but as we approach the  solar surface, the dynamics becomes even more complex.

The subsurface layers of the Sun, where the turbulent transport of heat toward the solar surface 
is done by means of robust fluid motions, and dynamical behavior of such layers are quite 
complex, very likely playing a significant role in the propagation of the magnetic field. In these 
layers, the heat coming from the solar interior is transported toward the solar surface by means 
of robust turbulent fluid motions, which we observed as patterns of convective motions, such as 
granulation, meso-granulation, and super-granulation. We could expect some type of chaotic 
motion, although more revealing is the existence of super-granular flows just below the 
photosphere. The analysis of long time series of observations from the Michelson Doppler 
Imager (SOHO) led to the discovery of super-granulation   traveling-wave patterns with periods of 
5--10 days and a lifetime of 2 days
\citep{art-Duvall1980,art-GizonDuvallSchou2003, art-GizonDuvallSchou2003er}. This result 
was later confirmed by other seismic measurements 
\citep{art-CorbardThompson2002,art-Schouetal2002}. The most recent measurement locates this 
flow just below the surface around $0.97 R_\odot$\citep[e.g.][]{art-Howe2009}.  

\smallskip
The analysis technique discussed in this work can be used to  
determine the impact of some of these flows, as well as to determine the contribution
of the turbulent convection for the formation of sunspots. Moreover,
it will permit us to quantify how these flows drive the magnetic variability 
during the past few centuries. Equally, this diagnostic tool can be used to validate 
the contribution of turbulent convection to the generation and evolution of the magnetic field
in the new generation of solar dynamo models~\citep{2009AnRFM..41..317M,rev-Charbonneau2010,20014Ilopes}.

\section{A new diagnostic of the solar magnetic cycle}
\label{sec-ndsmc}

Wavelet analysis is a powerful tool to probe specific features on large amounts of data. A 
particular wavelet transforms the signal to be analyzed into another representation, which 
displays the signal information in a more convenient form. In general, the wavelet function has 
the ability of decomposing the signal by means of two operations, by moving throughout the 
signal and/or by stretching or squeezing the signal. The choice of the wavelet to use depends on 
both the nature of our time series and on the type of features we are looking out for in the signal.  Our 
first concern on the choice of the wavelet to use is the reduction of the stochastic noise 
observed in the daily time-series data.
We choose to use a well-known continuous wavelet transform that is very well documented in the 
literature~\citep{book-mallat}: the Mexican hat wavelet transform. This wavelet is the second 
derivative of the Gaussian distribution, which, for the propose of this work, is written as
 \begin{eqnarray}
\psi(\sigma,t)=\frac{1}{4\sqrt{\pi \sigma}} 
\left(\frac{t^2}{2\sigma^2}-\frac{2}{\sigma}\right)
e^{-\frac{t^2}{4\sigma}}.
\end{eqnarray}
$\psi (\sigma, t) $ is the first function of the Gaussian family that satisfies the 
admissibility condition to be a wavelet \citep{book-mallat}. The parameterization of $\psi 
(\sigma, t) $ was chosen in such a way that the Fourier transform has a very simple expression. The 
Fourier transform of $\psi(\sigma,t)$,  which  we express as  $\widehat{\psi} (\sigma, \omega)$ is 
equal to  $-4\pi \omega^2 e^{-4\pi^2\omega^2 \sigma} $.

The two-dimensional wavelet transform  ${\cal W}(\tau,f)$  of the real data and
artificial data time series are shown in Figures~\ref{fig-Btotal} and~\ref{fig-vdpwavelet1}.
${\cal W}(\tau,f)$ is a function of $f$, a characteristic frequency, 
and  $\tau$ is the time-shift  parameter~\citep{book-mallat}.

A useful and intuitive quantity to show when studying periodic and non-periodic phenomena is 
the density power spectrum. In any type of transformation (including Fourier or wavelet 
analysis), the total energy of the signal is conserved. The Parseval theorem ensures us that the 
conservation of the total energy leads to the integral of the square of a signal over time being
equal to the integral of the square of its transform over the same time interval~\citep{book-mallat}.  
Therefore, the difference between two different transforms will be only due to differences 
between integrands. The ability to choose a specific wavelet is to identify the one transform 
that best shows the particular features that we are looking for. The power spectrum for a signal 
of length $T$ is given by
\begin{eqnarray}
P(f)=\frac{1}{T}\int_0^T E(\tau,f) d\tau,
\end{eqnarray}
where  $E(\tau,f)$ is the two-dimensional wavelet energy density function. $E(\tau,f)$ is equal
to $|{\cal W} (\tau,f)|^2/\pi$.  The power spectrum $P(f)$ is identical to the Fourier power 
spectrum (without the stochastic noise). The power spectrum computed using this expression
is presented in Figures~\ref{fig-Btotalpower},\ref{fig-vdpwavelet2} and~\ref{fig-OTwavelet}.

\section{The Observational Analysis of Sunspot Areas}
\label{sec-obs}

\begin{figure}[!htbp]
\centering
\includegraphics[height=6cm,width=8cm]{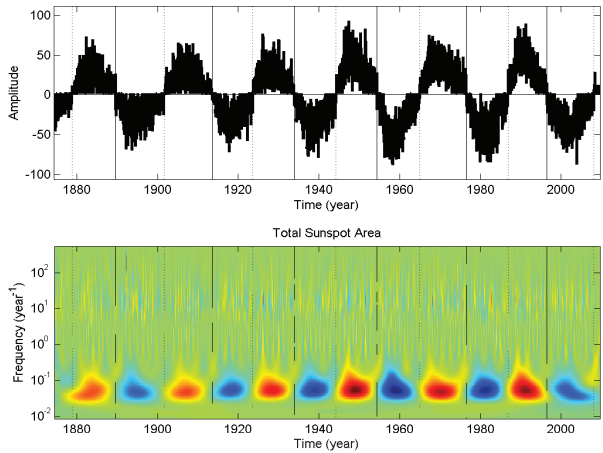}
\caption{Daily sunspot area time series: (a) The sunspot area time series in a Bracewell 
representation. This time series is proportional to the magnetic field (see text). (b) Wavelet 
transform of the daily sunspot area time series. Daily sunspot area data from 1874 to the present 
were compiled by NASA solar physics division. The vertical continuous and dotted lines define the 
dates of minimum of solar activity. The continuous lines  are the beginning of a full  magnetic 
cycle. The red and blue contour scales correspond to positive and negative variations 
of the magnetic field, respectively. The slight excess of intensity located at the right of each contour is a direct 
consequence of the difference between the rising and falling part of each demi-cycle.
This legend applies  to all the subsequent figures, unless stated otherwise.}
\label{fig-Btotal}
\end{figure}

\begin{figure}[!htbp]
\centering
\includegraphics[height=6cm,width=8cm]{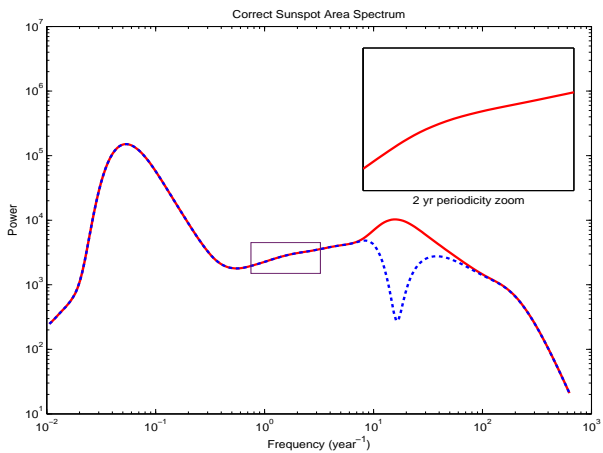}
\caption{Power spectral density of the sunspot area time series discussed in Figure~\ref{fig-Btotal}.
It is possible to identify two local maxima of intensity, occurring at two well-defined 
frequencies: one frequency with the value  $f_1\sim 0.054\;{\rm year}^{-1}$, the well-known 
sunspot cycle with a period of $18.43$ yr, and  a second frequency with the value $f_2\sim 16.11\;{\rm 
year}^{-1} $. This corresponds to a period of the order of $22.67\; {\rm day}$.
The blue dotted line corresponds to the  power spectral density of the sunspot area  where
the periodic source of $22.67\; {\rm day}$ was subtracted (see text).}
\label{fig-Btotalpower}
\end{figure}    

The  RGO has recorded sunspot areas using observational data 
from different observatories starting in 1874 until 1976. Following their work, the US Air Force 
(USAF) started compiling data from its own Solar Optical Observing Network, and 
more recently this work has been continued with the help of the US National Oceanic and 
Atmospheric Administration (NOAA), with much of the same information being organized and 
compiled until the present date. A complete reformatted data set has been made publicly 
available by the NASA/MSFC Solar Physics Division\footnote{Detailed information about the 
data can be found in \\ $http://solarscience.msfc.nasa.gov/greenwch.shtml$}
\citep{art-HathawayWilsonReichmann2002,art-HathawayWilson2004}. The available records 
are the daily sunspot area data measurement made between  1874 May 1 and 2009  June 30. The 
sunspot areas were measured in units of millionths of a hemisphere. The three time series analyzed 
in this work have each a total of 49,370 data points each. The daily data time series correspond to a global sunspot 
area, the northern hemisphere sunspot area, and southern hemisphere sunspot area. In this analysis, we 
have assumed that the magnetic field reverse occurs in the solar minimum.  We adopt the 
date of solar minimum activity,
\footnote{See the available table in \\ $ftp://ftp.ngdc.noaa.gov/STP/SOLAR\_ DATA$},  
the ones published in the publicly available data archives from  US NOAA. 
For convenience, in this work we choose the most recent minimum 
to be 2008 January.

The procedure for the estimation of the magnitude of magnetic field $B(t)$  using sunspot 
areas is quite similar to the one used for sunspot numbers.  It is worth noting 
that the correlation between the sunspot number and sunspot area is of $99\%$; therefore, there 
is no significant difference between these two proxies to estimate the intensity of the magnetic 
field \citep{art-HathawayWilson2004}. 
Since the magnetic field displays a 
period of approximately 22 yr with a polarity inversion every 11 yr,  we will use the 
Bracewell number~\citep{art-Bracewell1953} to explicitly  take this feature into account.  This 
quantity is defined as the square root of the sunspot area with a sign change at the beginning of 
each sunspot period. Therefore, each solar magnetic cycle will have one sunspot period with 
negative values and another with positive values. $B(t)$ can be computed as 
$B(t)=\beta_{A}\sqrt{A (t) }$, where $A (t) $ is the daily sunspot area measured. The 
parameter $\beta_{A}$ is a coefficient that converts the sunspot area proxy into the  magnitude of 
the  magnetic field. The studies of stability of tubes of magnetic flux suggest that the 
magnitude of the magnetic field inside the tachocline is between 1 and 10 T 
\citep{art-Magara2001,art-lp2009a}. A comparative representation of the intensity of the magnetic field at 
the surface and the sunspot area allows us to estimate the conversion factor $\beta_A$ 
\citep{art-RingnesJensen1960,rev-Solanki2003,art-pl2008a}. In this work we are interested in the 
variation of the mean magnetic field rather than its absolute value; therefore, without loss of 
generality, we choose $\beta_B$ to be equal to 1.  In Figure~\ref{fig-Btotal} we present the 
wavelet transform of the global sunspot area. The power density spectrum is shown in 
Figure~\ref{fig-Btotalpower}. It is possible to identify the long-term 
periodic solar magnetic field with a period 18.43 yr, among other features.
A detailed analysis of Figure~\ref{fig-Btotalpower} 
also shows that in the frequency interval from $0.5\;{\rm year}^{-1}$ until $10\;{\rm year}^{-1}$,
there are also other periodicities, like the biennial and Rieger periodicities.
However, as the amplitudes of such periodicities are much less intense than 
the ones previously mentioned, these are not clearly visible on the spectrum, although  
a tiny sinusoidal feature is visible in the mentioned region.
A discussion of the known periodic phenomena observed on the sunspot  data series,   
with periods varying from several months to several years, can be found in~\citet{2002A&A...394..701K}.

To check the validity of the density power spectrum
shown in Figure~\ref{fig-Btotalpower}, we have computed 
this using another type of wavelet.  
Interestingly, the general form of the density power spectrum 
was also found using the Morlet/Gabor wavelet. 
Equally, for this wavelet the leading periodicities  
have the same proportionality relation between their amplitudes,
a property consistent with the density power spectrum 
of the Mexican hat wavelet.

In the following section,
we propose a theoretical model to interpret the obtained density spectrum.   

\section{Model of the stochastic nature of sunspot areas}
\label{sec-smc}

\begin{figure}[!htbp]
\centering
\includegraphics[height=6cm,width=8cm]{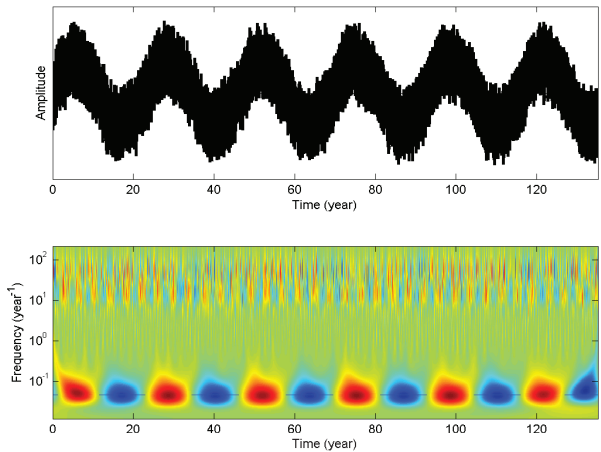}
\caption{Daily artificial data of the solar magnetic field $B(t)$.
It is accepted that $B(t)$ has two components (see text for details): $B_o(t)$,  a long-term scale 
variation  (Van der Pol-Duffing oscillator solution),  and $B_c(t)$, a  short-term scale variation  
(random uniform noise and wave-like perturbation). (a)  Six {\it solar magnetic} cycles  time 
series of a noise cyclical-like curve. (b) Contour plot of ${\cal W} (\tau,f)$ for the noise cyclical-
like curve in (a).}
\label{fig-vdpwavelet1}
\end{figure}

\begin{figure}[!htbp]
\centering
\includegraphics[height=6cm,width=8cm]{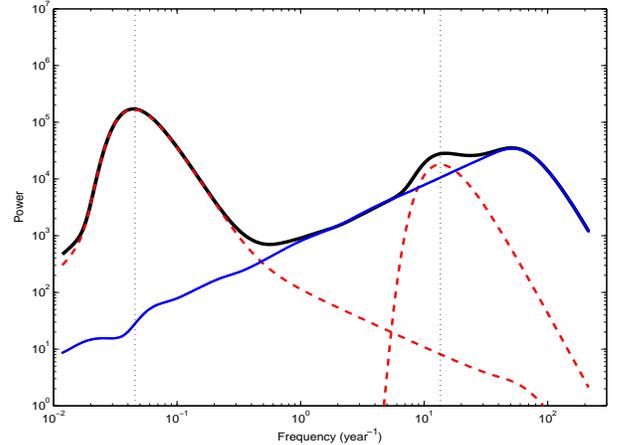}
\caption{Power spectral density of the noise cyclical-like curve discussed in Figure~\ref{fig-Btotal}. The 
vertical lines indicate the location of $f_o=\omega_o/2\pi$ (left line) and $f_p=\omega_p/2\pi$ (right 
line). This corresponds to the values 21.87  yr ($f_o\sim 0.04572 \; {\rm yr}^{-1}$) 
and 27 days ($f_p\sim 13.52 \;{\rm yr}^{-1}$).
The black curve corresponds to the full solution given by equation~(2). The blue curve 
corresponds to the random stochastic source (see text). The dashed red curves correspond to
the two single periodic sources with frequencies $f_o$ and $f_p$.}
\label{fig-vdpwavelet2}
\end{figure} 

The daily temporal evolution of the $B(t)$  on the Sun's surface can be interpreted as the final 
result of several physical mechanisms acting simultaneously on the continuous regeneration of 
the magnetic field. These magnetic field variations can be assumed to be constituted by two terms: 
one related to the long-term evolution of the solar magnetic cycle, as predicted by standard 
dynamo models, and a short-term component related to the excitation provided by the Sun's 
surface granulation and short-term periodic flows.
The turbulent flows preceding the formation 
of granules are not taken fully into account on the current solar dynamo models.  Therefore, the 
observed magnetic field on the Sun's surface, $B(t)$, will be the result of the superposition of  
these two distinct physical processes:   
\begin{eqnarray}
B (t)=B_o (t)+B_{c} (t),
\end{eqnarray}
where $B_o (t)$ and $B_{c} (t)$  are the long-term and the short-term components of evolution 
of the observed surface magnetic field, respectively. The component $B_o(t)$  is the long-term component of 
the magnetic field that is produced by the solar dynamo typically during a 22 yr period, and 
$B_{c}(t)$ is the short-term component of the magnetic field associated to the solar granulation 
and other short-term flows.

The evolution of the averaged magnetic field is described by a set of MHD
equations in a simplified form, known as the axisymmetric dynamo model 
\citep{rev-Charbonneau2010}. 
It has been shown that $B_o(t)$ corresponds to the temporal part of the 
 toroidal component of the averaged magnetic field~\citep{art-Mininni2001,art-pl2008b}. This 
axisymmetric formulation of the solar dynamo takes into account the contribution of the 
differential rotation and the meridional circulation flow. $B_o(t)$, which emerges at the solar 
surface, is the solution of a Van der Pol-Duffing oscillator\citep{art-lp2009a},
\begin{eqnarray}
\frac{d^2B_o}{dt}+\omega_o^2 B_o+\mu \left(3\xi B_o^2-1\right)\frac{dB_o}{dt}-\lambda 
B_o^{3}=0,
\label{eq-vpd}
\end{eqnarray}
where the coefficients $\omega_o$, $\mu$, $\xi$, and $\lambda$ are averages of the solar 
background used in the solar dynamo model. These quantities can be estimated from the 
observational data\citep{art-pl2008a, art-pl2008b}. Following our previous work and in 
agreement with what is published in the literature, we adopt the following set of averaged 
values for our numerical simulations\citep{art-lp2009a}:
$\omega_o=0.2873$, $\mu=0.1556$, $\xi=0.0221$, and $\lambda=1.5\times10^{-4}$. In the 
numerical simulation $B_o(t)$ takes values between -100 and +100 in arbitrary units.  The 
value of $\omega_o=2\pi f_o=0.2873$ corresponds to a magnetic cycle of a 22 yr period 
(i.e., a linear frequency $0.0457 {\rm yr}^{-1}$). 
This long-term periodicity is responsible for the strongest periodic source 
in the density power spectrum of the sunspot area data (see Figure~\ref{fig-Btotalpower}).

The short-term component, $B_c(t)$, emulates the stochastic excitation provided by 
small-scale flows, possibly like  the solar  granulation, 
including additional periodic flows (not included on the solar dynamo model), 
very likely related with the local hydrodynamics. The turbulent flows of the 
convection zone produce a complex pattern of structure observed on the Sun's surface. These 
motions seem to be arranging in hierarchy of surface outflows, which are the manifestations of 
the convective fluid motions below the photosphere. Usually these structures are characterized 
as patterns of granulations, meso-granulations and super-granulations. These well-defined 
granulation structures have typical scales of $1$, $10$ and $15-35$ Mm and averaged 
lifetimes of the $ 0.2$, $5$ and $24$ h, respectively. The daily variability observed in sunspot 
areas is very likely connected with the complex processes related with local flows 
and the chaotic motion of the short-scale magnetic fields associated with these granules.
This random magnetic behavior is very likely responsible for the stochastic nature of sunspot area
data (see Figure~\ref{fig-Btotalpower}). Therefore, we choose $B_c(t)$ to be  
a stochastic source, ${\cal A}_{r} U(t)$, where  ${\cal A}_r $ is the amplitude of a 
stochastic excitation source and  $U(t)$ is a uniform distribution that takes values  
between -1 and 1. 

Another source term that is identified in the sunspot area data is a periodic source, 
the second strongest peak, located at a higher frequency (see Figure~\ref{fig-Btotalpower}). 
There is the possibility that some of the periodic flows produced inside the star could 
present such a type of signature in the power spectrum. Unfortunately, 
other well-known processes could produce a similar structure.
One of our concerns relates to the possibility that such a periodic source is  
related to the synodic rotation, 
\footnote{The time for a fixed feature in the Sun to rotate to the same apparent 
position as viewed  from Earth. This chosen period roughly corresponds to  the rotation 
of sunspots located at a latitude of 26$^o$}, or Carrington rotation, assumed to be of the 
order of  27 days. Equally,~\citet{1992ApJ...397..337S} have identified a sunspot feature
with  a period of 26 days.

One of our concerns in the analysis of observational density power spectrum 
is to identify the origin of the 22.67 day periodicity.
Therefore, we choose to study the contribution of such a term to the
power spectrum.  Nevertheless,
regardless of the origin of this periodic source on the data, 
either being caused by the Carrington rotation or the 26 day  periodicity ~\citet{1992ApJ...397..337S},
or possibly produced by some other mechanism inside the star, 
we will study how the density power spectrum is modified by the presence
of such a periodic source. Accordingly, we choose our theoretical periodic 
phenomena as a fiducial period of 27 days.
 It follows that the second term of $B_c(t)$ is a periodic function 
with amplitude ${\cal A}_p $ and cyclic frequency $\omega_p$, hence, 
${\cal A}_p\sin{(\omega_p t)}$. Hence, 
\begin{eqnarray}
B_{c} (t)={\cal A}_{r} U(t)+{\cal A}_p\sin{(\omega_p t)}.
\end{eqnarray}
There are other periodicities in the observational density spectrum.
However, as these quantities have an amplitude much smaller than
the  one previously mentioned and the 22 yr solar cycle, they 
are not considered in this preliminary study.

Figure~\ref{fig-vdpwavelet1}(a) shows the $B(t)$ times series obtained for a simulation with the 
following fiducial values: ${\cal A}_r=100$ and  ${\cal A}_p=30$ in arbitrary units. The 
period of this source term is of 27 days, i.e., $\omega_p=85$.

The two-dimensional wavelet transform  ${\cal W}(\tau,f)$ of the artificial data time series $B(t)$ is shown in 
Figure~\ref{fig-vdpwavelet1}b. The oscillation pattern around $f\sim 10^{-1}\;{\rm year}$ is the distinct signature of the solar 
dynamo, i. e., $f_o=\omega_o/2\pi $. A careful analysis of Figure~\ref{fig-vdpwavelet1}b shows the 
second periodic signal near $10^2\;{\rm year}^{-1}$, i.e., $f_p=\omega_p/2\pi $.   

In Figure~\ref{fig-vdpwavelet2}, we shown the power spectrum of the artificial time series 
$B(t)$ computed previously (see Figure~\ref{fig-vdpwavelet1}).  The maximum of intensity for 
${\cal W}(\tau,f)$ and $P(f)$ occurs for values of $B(t)$, that have a cyclic behavior, namely, the 
two cyclic frequencies: $\omega_o$ and $\omega_p$. 
The random behavior of the time series, which we attribute to the stochastic nature of the sunspot areas, is  
shown in this power spectrum as a continuum, rather than the usual random noise observed in the  Fourier spectrum. 
The presence of a stochastic source term ${\cal A}_{r} U(t)$  increases 
significantly the power of the spectrum at high frequencies. More significantly, the amplitude 
power of the second cyclic source, ${\cal A}_p\sin{(\omega_p t)} $  increases with the 
magnitude of the stochastic source,  but it is still possible to identify the location of the maximum 
for this source term (compare the two curves in the power spectrum). 
The three source terms of the artificial time series are clearly identified in the power spectrum 
(see Figure~\ref{fig-vdpwavelet1}).
 
\begin{figure}[!htbp]
\centering
\includegraphics[height=6cm,width=9cm]{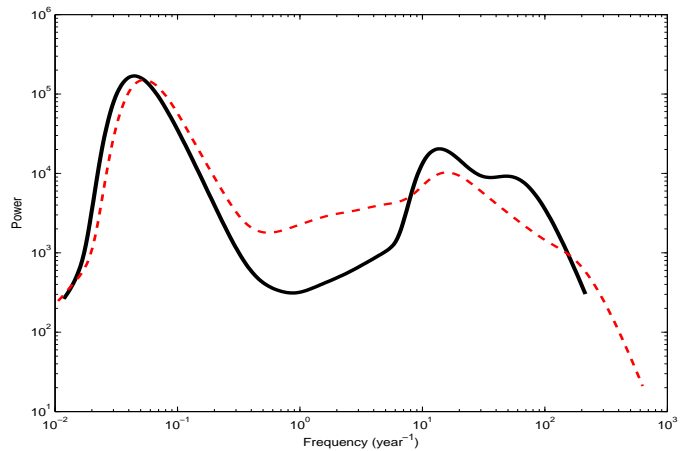}
\caption{Wavelet  transform of the daily sunspot area time series.
Comparison between the spectra of the total sunspot area observed data and the
sunspot proposed model. The black curve is computed for the theoretical model
with two periodic functions, the periods of 21.87 year and 27 days.
The dashed red line corresponds to the sunspot data,
where  local maximums occur for the periods 
$18.43\; {\rm yr}$ and  $22.67\; {\rm days}$.}
\label{fig-OTwavelet}
\end{figure} 
 
\section{Discussion} 
\label{sec-discussion}
 The daily time series displays the inherent noisiness of the solar cycle, very likely related to the 
turbulent flows of the Sun's surface. The data series changes widely daily. The Sun rotates with 
a period of 25 days (at the equator), during which the solar activity can change significantly,
which corresponds to a synodic rotation period of 26.24 days, although the 
Carrington rotation is assumed to be of the order of 27 days. Earth 
observers can only measure sunspots from the observed hemisphere. To compensate this fact, the 
usual procedure is to use some kind of temporal filter to smooth the data, typically a 13-month 
running mean. This procedure strongly reduces  the stochastic noise at high frequencies. 
Unfortunately, this procedure also removes the contribution of perturbations to the sunspot 
cycle, with high frequencies.  By using the wavelet procedure previously introduced, we have 
analyzed the time series over all the entire range of frequencies allowed. Using the proposed 
technique, there is no need for mean averages, once the full time series can be completely reproduced  
(in the mathematical sense)  as a sum of basic time series at different scales 
(in our case, frequencies). Each scale is somehow an average of the observational time series at that 
scale. More importantly, the small scales are maintained on the series. In fact, the study of small 
scale processes in the Sun can be done by this method, once we can assume that the 
physical processes occurring at a small-scale, which are measured in the observed hemisphere, also 
occur in other parts of the Sun. The weakness of such a procedure is that. in principle,  a periodicity
related to the Carrington rotation should appear, if not cancelled out by some average behavior.
We believe that the total daily areas of the observed Sun's globe or each of its hemispheres,
on average, are not affected by the Carrington rotation period, once this contribution in average  
cancels out.

\begin{figure}[!htbp]
\centering
\includegraphics[height=6cm,width=9cm]{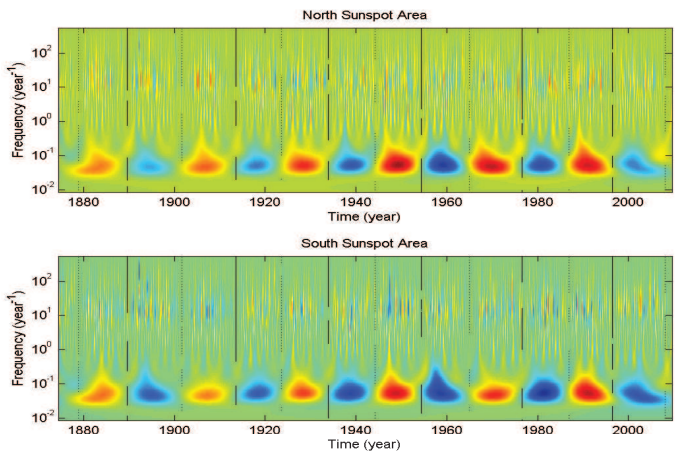}
\caption{Wavelet  transform of the daily sunspot area time series of  the northern and southern hemisphere.
In both  hemispheres we find the same typical behavior, namely, the existence of an excess of energy 
near   a frequency of the order of $10^2\; {\rm year^{-1}}$. For both hemispheres,
it is possible to identify a small excess of intensity for lower frequencies.}
\label{fig-NSwavelet}
\end{figure}

\begin{figure}[!htbp]
\centering
\includegraphics[height=6cm,width=8cm]{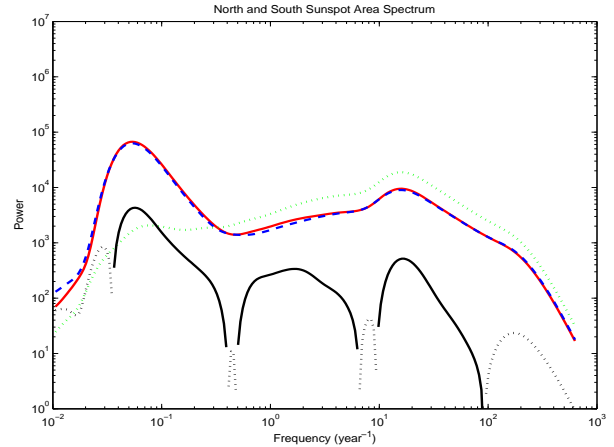}
\caption{Power spectral density of  the sunspot area time series  of the northern and southern 
hemispheres as  discussed in Figure~\ref{fig-NSwavelet}. The continuous red curve corresponds to the power spectral density
of the northern hemisphere and the  dashed blue curve to power spectral density of the southern hemisphere. 
The continuous-dotted black curve shows the spectral regions where the northern hemisphere is more active
than the southern hemisphere (continuous) or the reverse (dotted line). 
The dotted green line corresponds to the power spectral density computed from wavelet intensity 
differences between  the two hemispheres. The first peak ($f_1$) shows a strong correlation between hemispheres, 
whereas the second peak ($f_2$) shows no correlation at all between the two hemispheres. }
\label{fig-NSwaveletpower}
\end{figure} 

\subsection{The global sunspot area}

The plot of the magnetic field computed from the global sunspot area is shown in 
Figure~\ref{fig-Btotal}, and the power spectrum is shown in Figure~\ref{fig-Btotalpower}. We 
distinguish three distinct components related to the solar magnetic field: one spectral peak 
produced by an excess of power is located near $f_1\sim 	 0.054\; {\rm year}^{1}$, a second  
spectral peak is located near $f_2\sim 16.11\; {\rm year^{-1}}$, and a third component is the 
increase of power toward the high frequencies, which we associated with the daily
turbulent excitation of sunspots. 

The forms of these components are similar to the ones discussed in the artificial data time series 
(see Figures~\ref{fig-vdpwavelet1} and~\ref{fig-vdpwavelet2}). 
In Figure~\ref{fig-OTwavelet} we compare the theoretical model computed (discussed in the
previous section) with the observed power spectrum. The first striking result is the similarity between the two curves.

The strongest excess of power occurs for the frequency $f_1\sim 0.054\; {\rm yr}^{-1}$, 
or 18.42-year  period. Although the source is cyclic, it presents quite substantial intensity variations 
from  cycle to cycle. The contour plot of ${\cal W}(\tau,f)$ is asymmetric, presenting an excess of 
intensity on the right side, similar to the theoretical model previously discussed. 
The second excess of power of the time series is located at a frequency of 
$16.11\; {\rm year^{-1}}$   which corresponds to an apparent period of 22.67 days. 
It is worth mentioning 
that such a secondary peak on the power density spectrum has an intensity value only 15\% smaller  
than the main peak associated with the 22 yr solar magnetic cycle. It seems to be a quite strong source.
In Figure~\ref{fig-Btotalpower} we show the shown the case where this last periodic term is removed 
 and the overall spectrum maintains the same structure.

We have  made other analyses of these time series, namely, we have computed the
power spectrum from the original sunspot area data time series, without the 
introduction of the Bracewell numbers, and  we have found the same excess of power in 
the same locations of the power spectrum.  Once again, this is observed in all the three time series,
observed Sun's disk, northern hemisphere, and southern hemisphere. 
These results confirm our initial hypothesis that there is an oscillation component of 22.67 days  
on the sunspot time series.  

However, establishing the exact physical origin of such periodic source is more difficult. 
It could be related to the Carrington rotation, the 26 day periodicity found by~\citet{1992ApJ...397..337S},
or it could have originated in the interior of the Sun.
Nevertheless, the 22.67 day period is quite different from  
these other two periodicities.
To investigate the origin of such a secondary source, we have made a spectral
analysis of each of the solar magnetic cycles. We found that the secondary
source peaks in each of the solar magnetic cycles continues to exist, but the period 
takes values between 21.5 and 24 days. These values occur between different solar magnetic cycles, 
but also between different hemispheres, even if values near  22.67 days occur with more frequency.
This is the stronger indicator that suggests that  the origin of such a secondary source is related
to physical processes inside the Sun.
Actually, there are several internal flows that operate in similar timescales.
The internal rotation  and additional averaged flows could be at the origin of the 
formation of  observed sunspot patterns \citep{art-Wolff1995,art-GizonDuvallSchou2003,art-GizonDuvallSchou2003er,art-Howeetal2006}. 
The available data do not allow us to test further the hypothesis that
this power access at high frequency have originated by the Carrington rotation. 
Finally, it is 
possible to identify an almost  linear increase of power between frequencies $0.5\; {\rm yr^{-
1}}$ and $5\; {\rm yr^{-1}}$ . This very likely is related to turbulent noise of granulation. 
This stochastic noise is present all over the spectrum, but it is more noticeable in the mentioned 
frequency interval (see Figure~\ref{fig-Btotalpower}). In Figure~\ref{fig-OTwavelet} 
the similarity of behaviors between the theoretical model and observation is very clear, although  there is scope for
improvement. 

\subsection{The North and South sunspot area time series}

The comparative analyses of the power spectrum for both hemispheres 
(see Figures~\ref{fig-NSwavelet} and~\ref{fig-NSwaveletpower}) show very similar results. 
There is an indication of an asymmetric behavior between the northern and the southern hemisphere.
For very low  frequencies ($f \le 3 10^{-2}\;{\rm yr}^{-1}$), 
the southern hemisphere seems to present an excess of energy when compared 
with the northern hemisphere (see Figure~\ref{fig-NSwaveletpower}), as has been reported 
in previous  works\citep{art-HathawayWilson2004,2007A&A...475L..33D}.
Several authors have found evidence of this asymmetry for different proxies at distinct timescales~\cite[e.g.,][]{1996SoPh..167..409V,2008AdSpR..41..297V,2009SoPh..255..289L,2013ApJ...768..188C}.
However, for larger frequencies  ($f \ge 3 10^{-2}\;{\rm yr}^{-1}$) the situation is quite the reverse:
the northern hemisphere seems to present an excess of energy in relation to the southern hemisphere.
In this spectral interval, we found three strong excitation sources, 
which present quite different  characteristics. There is a systematic excess of power for the 
northern hemisphere when compared with the southern hemisphere, namely, near the two
studied periodic sources, as well as in the intermediate region where the biennial and Rieger periodicities are known to be 
located. This is also the region of the spectrum where the turbulent processes of the convection zone seem to 
make the most important contribution for the dynamics of sunspots.

The source related with the 18.42 yr period associated with the long-term solar magnetic cycle 
is strongly correlated between the two  hemispheres, although the source located at a period of 22.67 days  
is not correlated  (see Figure~\ref{fig-NSwaveletpower}).   This is clearly illustrated by computing 
the power spectral density of the  difference between the two hemispheres, where for the low frequency 
the effects cancel out, and  for the high frequency they add up,  i.e., the 18.42 yr period physical process 
is strongly correlated,  but the 22.67 day period seems to be equally independent of each hemisphere. 
This suggests that the 22.67 day periodicity has a phenomena strongly asymmetric between the
two hemispheres. Moreover, as per Figure~\ref{fig-NSwaveletpower}, the physical process is more
pronounced in the northern hemisphere.

\subsection{Comparison with Lomb-Scargle spectral analysis}

We complement the wavelet study of the total sunspot area with a Lomb-Scargle (LS) spectral analysis. This technique was developed for interrupted data sets in astrophysics~\citep{art-Lomb1976,art-Scargle1982} and is often used to do spectral studies in different sciences. Actually, the total daily area of sunspots is an uneven time series because of the days when there were no observations (represented in the data sets as -1). LS spectra provide the significant frequencies (in a statistical sense) and their respective amplitudes, enabling a proper evaluation of the dominant periods that influence the data. The program used in the present study is an LS implementation\footnote{http://www.mathworks.com/matlabcentral/fileexchange/993-lombscargle-m 
(retrieved on 2014 January 11)}~\citep{1992nrca.book.....P}.

\begin{figure}[!htbp]
\centering
\includegraphics[height=6cm,width=8cm]{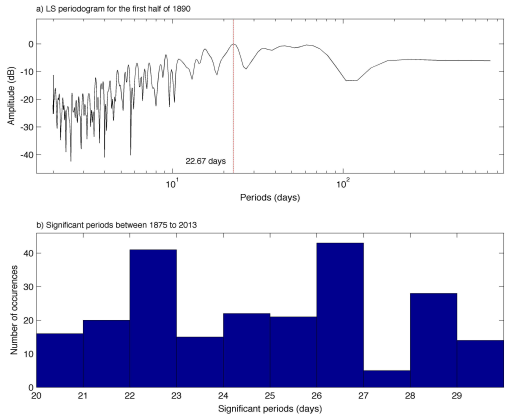} 
\caption{Upper panel: representative LS periodogram for the first half of 1890. 
Lower panel: histogram of the significant frequencies in the 20--30 days range.
The histogram shows two periodicities located at 22.67 and 26.6 days (see main text).
 }
\label{fig:8}
\end{figure}

The parameters used are hifac = 1 (which defines the frequency limit as hifac times the average Nyquist frequency) and ofac = 4 (oversampling factor). Moreover, to verify the statistical significance of the 22.67 day period found above, we calculated the LS spectrum for half-year range from 1875 until 2013 (this means 276 LS spectra).
From them we select the significant periods in the range of 20--30 days; some half-years do not have periods in this range, the majority have one, others have two significant periods, and only one has three. In Figure~\ref{fig:8}, the upper panel shows the LS periodogram for the first half of 1890, where the 22.67 day period is really clear. The lower panel represents the values of the significant periods found in a histogram. It is important to mention that in a half-year it is possible to have six to eight complete cycles with periods in the range of 20-30 days, enough to be detected if present. Two peaks dominate the histogram: one for periods between 22 nd 23 days with 41 occurrences, the other for periods between 26 and 27 days having 43 counts.  This is significant because the first peak corresponds to the 22.67 day periodicity.
The second peak corresponds to either the Carrington rotation of the 25.8 days periodicity found by~\citep[]{1992ApJ...397..337S}, although the second interpretation seems to be more reliable.     
The clear identification of such periodicity clearly shows that the first periodicity
is distinct from the second one. 

\begin{figure}[!htbp]
\centering
\includegraphics[height=6cm,width=8cm]{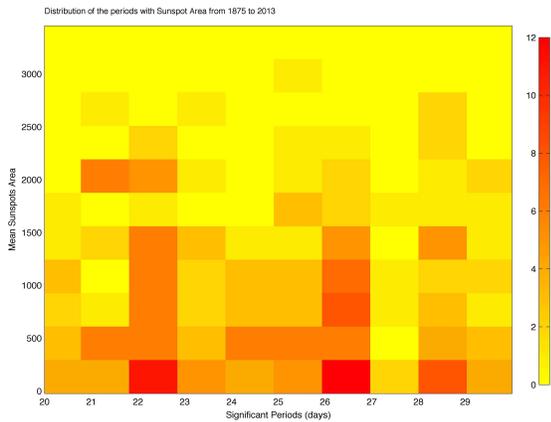} 
\caption{Periodicity relations with the sunspot areas: 
2D histogram distribution of periods in the 20--30 day range as a function of the half-year mean sunspot areas.}
\label{fig:9}
\end{figure}

\begin{figure}[!htbp]
\centering
\includegraphics[height=6cm,width=8cm]{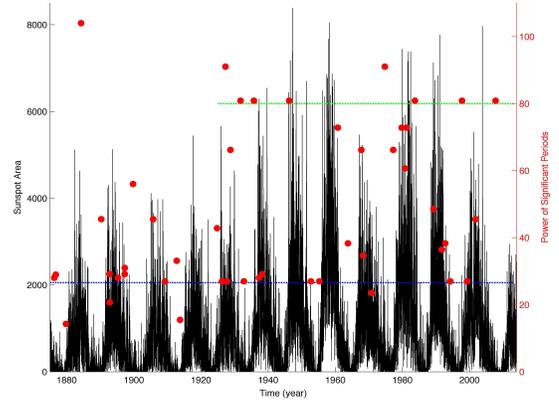} 
\caption{Horizontal lines define the two typical power scales:
one around 28 (marked as a blue dotted line) and another around 80 (marked as a green dotted line).  
}
\label{fig:10}
\end{figure}

To clarify the dependency of the significant periods on the evolution of the sunspot area, we calculate the mean value of the sunspot area for every half-year similarly to what he have done for the LS periodograms. After that, we plot the distribution of the significant periods in the 20--30 day range as a function of the half-year mean sunspot area using an efficient 2D histogram, Figure~\ref{fig:9}. The dominance of the periods in the range of 22--23 days and 26--27 days is clear; this is consistent with the 1D histogram (see Figure~\ref{fig:8}). Naturally, there are fewer counts for higher sunspot areas; this is because the occurrence of sunspot areas is progressively decreases as the areas increase, and not because they do not generate periodicities in the range of interest.
Moreover, the occurrence of periods in these two ranges seems not to be dependent on the sunspot area. 
Even though no direct relation exists between the 22 and 23 day periodicities and the solar cycle strength,
there is some indication of a possible connection. To illustrate this point, 
we represent the power of the significant periods in the range of 22--23 days 
along with the sunspot area from 1874 until 2013. Figure~\ref{fig:10} shows that power clusters 
around two distinct levels, one around 28 and another around 80. It is clear  that the power of $\sim$80 became more frequent after 1930, and this could be tentatively attributed to the onset of the modern maximum from 1950 up to 2000.

\

\section{Conclusion}
\label{sec-Conclusion}

In this work, we made a spectral analysis of the daily sunspot areas, to identify the different 
mechanisms underlying the evolution of the magnetic field, or to be more exact regarding the dynamics 
of the surface's magnetic field. Our study of the sunspot area time series, revealed clear 
signatures pointing to the possibility that the global behavior of the solar magnetic cycle is 
affected by different physical processes. 

The leading result of this work is found in Figure~\ref{fig-OTwavelet}, where 
we compare the power  spectral density of the theoretical model 
with the equivalent spectrum computed from daily time series of sunspot areas.
In both spectra it is possible to identify three well-defined structures:  
two major periodic sources located at 18 yr and 23 days and a spectral continuum 
in the power spectral density, clearly visible between these two leading periodicities.
In between these periodicities, the intensity of the spectral continuum increases as the frequency decreases.
Moreover, the spectral line widths of the peaks corresponding to 
the 18 yr and 23 days periodic phenomena somehow measure the variability of the 
periods of such phenomena during the duration of the time series. 
 
The 18 yr periodicity is the well-known self-regulated mechanism of the 
dynamo that converts kinetic energy into magnetic energy and is responsible for the 22 yr solar 
magnetic activity, which we estimate to be of the order of 18.42 yr. The second term seems to be  
intrinsically of stochastic nature, possibly  associated with the turbulent processes of the 
convective region and granulation. Finally, this  method of analysis also reveals another contributor 
to the evolution of the magnetic field: a cycle source with a period of 22.67 days. 

An important result of this study is the identification of the spectral continuum in the sunspot area time series, 
mostly visible as a linear continuum between the 18 yr and 23 day periodic sources. 
This continuum is a signature of the contribution of granulation and 
turbulent convection for the variability of the sunspot areas.
It is worth recording that  the magnetic field  on the solar surface displays a systematic pattern 
of activity that is mapped by sunspot numbers and areas. 
These sunspots are produced by localized concentration of magnetic fields 
with intensities of the order of a kilogauss. As the sunspots inhibit the convection 
in the Sun, they change significantly the global behavior of turbulence and magnetism in the solar convection zone.
In particular, as the intensity of the magnetic field within a nearly formed  sunspot decreases 
with time,  the impact of  granulation on the sunspot area becomes more relevant. 
Accordingly, by measuring the random behavior of sunspot areas over time, we are obtaining  
information about the impact of granulation on the evolution of the magnetic field.

Finally, we have been able to reproduce most of the results found in
the observational data by using a simple solar oscillator model. 
We have found that an oscillator model with a periodic source of 27 days (which could be either be the Carrington rotation 
or the 26 days periodicity~\citet{1992ApJ...397..337S}) presents a very similar structure to the one found 
in the observed power density spectrum.  Nevertheless, the value determined from observational data, i.e.  22.67 days, 
is quite different from these two periodicities. Moreover, we have made an independent
LS spectral analysis of the sunspot data and have confirmed that
the 22.67 day periodicity is distinct from the two others previously mentioned.
Nevertheless, we found that this periodicity is located near other periodicities.    
There is the strong possibility that such specific periodic structure is produced 
by some unknown internal process inside the convection zone.
The analysis  of this source term in each of the hemispheres indicates that term has a very identical nature in 
both hemispheres, i.e., there is no correlation between the two hemispheres. This is quite different from 
the 18.42 yr periodic term, which has strong correlation between the hemispheres. 
The asymmetry between hemispheres is stronger for the higher frequencies, especially 
near the two periodic sources.

\acknowledgments
The authors thank the referee for the thorough 
and detailed review of the manuscript, as well as for the bibliography suggestions 
that have enriched its contents, making it of interest to a more eclectic audience.
The authors also thank Elisa Cardoso for her contribution to this work. 
The work of I.L. was supported by grants from "Funda\c c\~ao para a Ci\^encia e Tecnologia"  
and "Funda\c c\~ao Calouste Gulbenkian". 
H.G.S. acknowledges the support of two Portuguese institutions: the Science and Technology 
Foundation (FCT) for the Post-Doc grant SFRH/BPD/63880/2009 and the Calouste Gulbenkian 
Foundation for the award "Est\'{\i}mulo \'{a} Criatividade e \'{a} Qualidade na Actividade de 
Investiga\c c\~ao" in the Science program of 2010. H.G.S. is also grateful to Mourad Bezzeghoud. 
We would like to acknowledge the Solar Physics Team from the NASA/MSFC 
for making  available the solar data files for public use.

\bibliographystyle{mn2best}

\end{document}